\documentclass[twocolumn,prl,superscriptaddress,showpacs,showkeys,nofootinbib]{revtex4-1}

\usepackage[english]{babel}
\usepackage{amsmath,amssymb,graphicx}
\usepackage{algorithmic}
\usepackage{enumerate}
\usepackage{times}
\usepackage{color}
\usepackage{soul}
\usepackage{textcomp}
\usepackage[pdftex]{hyperref}
\hypersetup{colorlinks=true,linkcolor=blue,citecolor=blue,urlcolor=blue}

\newcommand{\X}{{\mathbf{x}}}
\newcommand{\Z}{{\mathbf{z}}}
\newcommand{\U}{{\mathit{U}}}
\newcommand{\D}{{\mathit{D}}}
\newcommand{\N}{{\mathit{N}}}
\newcommand{\V}{{\mathit{V}}}
\newcommand{\J}{{\mathit{J}}}
\newcommand{\n}{{\mathit{n}}}
\newcommand{\m}{{\mathit{m}}}
\newcommand{\kk}{{\mathit{k}}}
\newcommand{\p}{{\mathit{p}}}
\newcommand{\lv}{{\mathit{l}}}

\begin{document}

\title{Uncertain fate of fair sampling in quantum annealing}

\author{Mario S. K\"onz}
\email[]{mkoenz@itp.phys.ethz.ch}
\affiliation{Institute for Theoretical Physics, ETH Zurich, 8093 Zurich,
Switzerland}

\author{Guglielmo Mazzola}
\affiliation{Institute for Theoretical Physics, ETH Zurich, 8093 Zurich,
Switzerland}

\author{Andrew J. Ochoa}
\affiliation{Department of Physics and Astronomy, Texas A\&M University,
College Station, Texas 77843-4242, USA}

\author{Helmut G. Katzgraber}
\affiliation{Department of Physics and Astronomy, Texas A\&M University,
College Station, Texas 77843-4242, USA}
\affiliation{1QB Information Technologies (1QBit), Vancouver,
British Columbia, Canada V6B 4W4}
\affiliation{Santa Fe Institute, 1399 Hyde Park Road, Santa Fe,
New Mexico 87501 USA}

\author{Matthias Troyer}
\affiliation{Institute for Theoretical Physics, ETH Zurich, 8093 Zurich,
Switzerland}
\affiliation{Microsoft Quantum, Microsoft, Redmond, Washington 98052, USA}

\date{\today}
\begin{abstract}

Recently, it was demonstrated both theoretically and experimentally on
the D-Wave quantum annealer that transverse-field quantum annealing does
not find all ground states with equal probability. In particular, it was
proposed that more complex driver Hamiltonians beyond transverse fields
might mitigate this shortcoming. Here, we investigate the mechanisms of
(un)fair sampling in quantum annealing.  While higher-order terms can
improve the sampling for selected small problems, we present multiple
counterexamples where driver Hamiltonians that go beyond transverse
fields do not remove the sampling bias. Using perturbation theory we
explain why this is the case. In addition, we present large-scale
quantum Monte Carlo simulations for spin glasses with known degeneracy
in two space dimensions and demonstrate that the fair-sampling
performance of quadratic driver terms is comparable to standard
transverse-field drivers.  Our results suggest that quantum annealing
machines are not well suited for sampling applications, unless
post-processing techniques to improve the sampling are applied.

\end{abstract}


\maketitle

Quantum annealing (QA)
\cite{finnila:94,kadowaki:98,brooke:99,farhi:01,santoro:02,das:05,santoro:06,das:08,morita:08}
is a heuristic designed to harness the advantages of quantum mechanics
to solve optimization problems. The performance of QA and, in
particular, QA machines such as the D-Wave Systems Inc.~devices are
controversial to date
\cite{dickson:13,pudenz:13,smith:13,boixo:13a,albash:15a,ronnow:14a,katzgraber:14,lanting:14,santra:14,shin:14,vinci:14,boixo:14,albash:15,albash:15a,katzgraber:15,martin-mayor:15a,pudenz:15,hen:15a,venturelli:15a,vinci:15,zhu:16,mandra:16b,mandra:17a,mandra:17y}.
Most studies have focused on finding the minimum value of a binary
quadratic cost function (problem Hamiltonian), yet less on the {\em
variety} of solutions obtained when repeating the optimization procedure
multiple times. Important applications that rely on sampling, such as
satisfiability (SAT)-based probabilistic membership filters
\cite{weaver:14,schaefer:78,douglass:15,herr:17}, propositional model
counting and related problems \cite{jerrum:86,gomes:08,gopalan:11}, or
machine learning \cite{hinton:02,eslami:14} rely on ideally uncorrelated
states. This sought-after {\em fair sampling} ability of an algorithm,
i.e., the ability to find (ideally all) states associated with a cost
function with (ideally) the same probability, is thus of importance for
a variety of applications. Moreover, the ability of an algorithm to
sample ground states with similar probability is directly related its
ergodicity which strongly influences the efficiency of optimization and
sampling techniques.

Following small-scale studies \cite{matsuda:09}, Ref.~\cite{mandra:17}
recently performed systematic experiments on the D-Wave 2X annealer. The
results demonstrated that quantum annealers using a transverse-field
driver are biased samplers, an effect also observed in previous studies
\cite{boixo:13a,albash:15a,king:16}. Matsuda {\em et al.}
\cite{matsuda:09}~conjectured that more complex drivers might alleviate
this bias, something we test in this Rapid Communication.

Binary optimization problems can be mapped onto $\mathit{k}$-local spin
Hamiltonians.
Without loss of generality
we study problem
Hamiltonians with $\N$ degrees of freedom in a $\Z$-basis of the form
\begin{equation}
{\cal H}_{\rm P} = -\sum_{i,j = 1}^N J_{ij} \sigma^\Z_i \sigma^\Z_j,
\end{equation}
where $\sigma_i^\Z$ is the $\Z$-component of the Pauli operator acting
on site $\mathit{i}$. Note that local biases can also act on the variables.  For
such a problem Hamiltonian, in principle, a driver of the form
\begin{equation}
{\cal H}_{\X,\n} = \sum_{i=1}^\n \Gamma^{\X,i}[\otimes \sigma^\X]^i
\end{equation}
would induce transitions between all states if $\n$ is set to $\N$ and therefore ensure a fair
sampling, provided the anneal is performed slow enough.  Unfortunately,
such a driver is hard to engineer and, at best, one can expect
drivers of the form ${\cal H}_{\X,2} = -\sum_j \Gamma^\X \sigma_j^\X +
\sum_{j,k} K_{j,k}^\X\sigma_j^\X \sigma_k^\X$.
Quantum fluctuations are induced by
the driver and then reduced to sample states from the
problem Hamiltonian, i.e., ${\cal H}(\mathit{t}) = (1-\mathit{t}/\mathit{T}) {\cal H}_{\X,n} +
(\mathit{t}/\mathit{T}) {\cal H}_{\rm P}$, where $\mathit{t} \in [0,\mathit{T}]$, $T$ the annealing time,
and $\mathit{n}$ the order of the interactions in the driver. For an
infinitely-slow anneal, the adiabatic theorem
\cite{kadowaki:98,farhi:00} ensures that for $\mathit{t} = \mathit{T}$ a (ground)
state of the problem Hamiltonian is reached. It is therefore desirable
to know if after an infinite amount of repetitions, the process results
in {\em all} minimizing states, i.e., fair sampling.

\begin{figure}[h!]
\includegraphics[width=1\columnwidth]{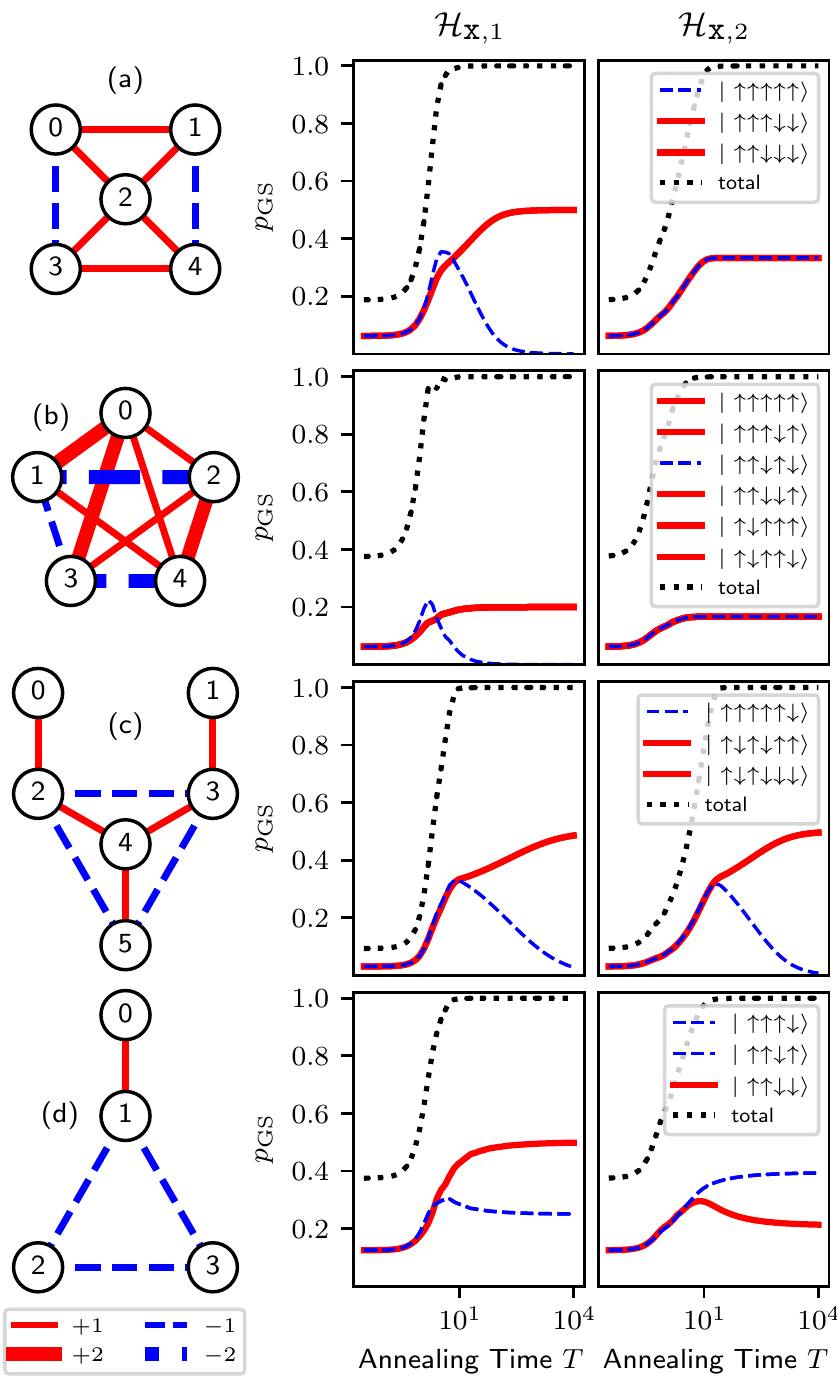}
\caption{\label{Fig1}
Toy problems with up to $\N = 6$ variables and both
ferromagnetic (solid lines) and antiferromagnetic (dashed lines)
interactions of different strength (thickness of the lines) integrated
using \textsc{qutip} \cite{johansson:13}. Data for both
transverse field (${\cal H}_{\X,1}$) and quadratic  (${\cal H}_{\X,2}$)
drivers are shown. The data show the instantaneous probability $\p_{\rm
GS}$ to find different states that minimize the cost function (up to
spin-reversal symmetry) as a function of anneal time $\mathit{t}$.  In all cases,
the first spin (labeled with $0$) is in the state $|\uparrow\rangle$.
(a) Toy problem studied in Ref.~\cite{matsuda:09} where drivers with $\n
= 2$ sample the states fairly. (b) Similar behavior to (a),
however, the unfair sampling sets in earlier in the anneal.  (c) Even
the inclusion of $\n = 2$ drivers does not remove the unfair sampling.
As in (a) and (b) at least one state is suppressed.  Note
that a driver Hamiltonian with $\n = 4$ results in fair sampling.  (d)
The sampling is not exponentially biased. However, one state occurs
twice as likely as the others. Note that the sampling probabilities swap
when going from $\n = 1$ to $\n = 2$.}
\end{figure}

Here we analyze the behavior of more complex drivers of the form ${\cal
H}_{\X,n}$ ($\mathit{n} > 1$) on the fair sampling abilities of QA.  Following
Ref.~\cite{matsuda:09} we first study small systems where the
Schr\"odinger equation can be integrated using \textsc{qutip}
\cite{johansson:13}. We have exhaustively analyzed all possible graphs
with up to $\N = 6$ with both ferromagnetic and antiferromagnetic
interactions and show in Fig.~\ref{Fig1} paradigmatic examples that
illustrate different scenarios using drivers with $\mathit{n} \le 2$.  Even for
some of these small instances, in some cases the inclusion of
higher-order driver terms does not remove the bias. If we anneal
adiabatically, i.e., $\mathit{T}$ large enough, the instantaneous ground states
are never left, which means towards the end of annealing at
$\mathit{T}-\mathit{\lambda}$ (for a small $\mathit{\lambda} > 0$) the system is in the ground state of
${\cal H}(\mathit{T}-\mathit{\lambda})$. This observation is key to predicting the
sampling probabilities for different degenerate ground states. These
probabilities are given by squaring the amplitudes of the lowest
eigenvector of ${{\cal H}(\mathit{T}-\mathit{\lambda})}$, assuming for now the small
contribution from the driver lifts the degeneracies. Because  ${\cal
H}(\mathit{T}-\mathit{\lambda})$ can be viewed as ${\cal H}_{\rm P}$ perturbed by
${\cal H}_{\X,n}$, we analyze fair sampling using a perturbative
approach \cite{lanting:17}. To better quantify the fair-sampling
behavior of a given system, we use the term {\em hard
suppression} (i.e., total suppression) if the sampling probability is
$0$ for a particular ground-state configuration at the end of the anneal
and the term {\em soft suppression} if a particular state is
undersampled by a certain finite fraction in comparison to other
minimizing configurations.  Finally, we complement these studies with
quantum Monte Carlo simulations for large two-dimensional Ising
spin-glass problems following Ref.~\cite{mandra:17} and discuss the
effects of higher-order drivers. Our results show that QA is not well
suited for sampling applications, unless post processing techniques are
implemented \cite{ochoa:18x}.

\begin{figure}[h!]
\includegraphics[width=1\columnwidth]{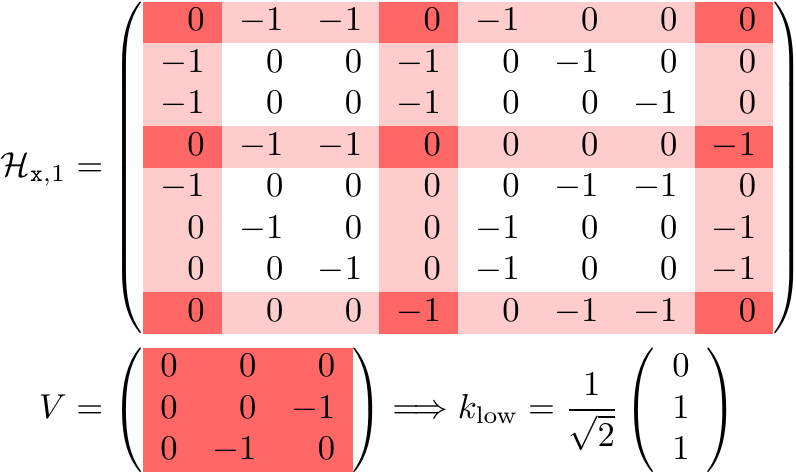}
\caption{\label{FigVD}
To obtain the sampling probabilities, the ground-state eigenvectors
$|g_i\rangle$ need to be known, represented here as shaded rows and
columns in the matrix ${\cal H}_{\texttt{x},1}$, since the solution of
the diagonal ${\cal H}_{\rm P}$ is a classical one. One then needs
to analyze the subspace $\V$ that is formed by restricting the driver (here ${\cal
H}_{\texttt{x},1}$) to space spanned by the ground states of ${\cal H}_{\rm P}$. The lowest
eigenvector(s) determine the sampling probabilities. In this example,
there is one lowest eigenvector and the first ground state corresponding
to the top column (first row) is suppressed (all spins up) and will never
be sampled in an adiabatic anneal.}
\end{figure}

{\em Perturbation theory}.
In the following, we show how to determine the sampling probabilities, as
well as the influence the driver has on it. Similar work has been done in \cite{siberer:18}. In short, if we apply ${\cal H}_{\rm D}$
as a perturbation of strength $\mathit{\lambda}$ to ${\cal H}_{\rm P}$, some degeneracies
will be lifted, i.e. the perturbed ground-state space is smaller. The ground-state space
is never left during an adiabatic anneal, hence it will not be possible to reach the
entire ground-state space of the unperturbed hamiltonian by annealing in the generic case.
This analysis hold for any driver hamiltonian ${\cal H}_{\rm D}$, not
just the stoquastic ${\cal H}_{\X,n}$-type drivers we use in this work.
In non-degenerate perturbation theory, the first order corrected wave
function $|\n \rangle$ is given by
$|\n \rangle = |\n^0 \rangle + \lambda\sum_{\m\neq \n}\frac{\langle \m^0|H_{\rm D}|\n^0\rangle}{E_\m^0-E_\n^0}|\m^0\rangle$,
where $|\n^0 \rangle$ are the eigenstates and $E_\n^0$ the eigenvalues of the unperturbed
hamiltonian ${\cal H}_{\rm P}$. If states $\m\neq \n$ are degenerate, i.e. $E_\m^0=E_\n^0$, there is a singularity.
To avoid it, degenerate perturbation theory is requires linear combinations
$| \alpha^0\rangle$ which satisfy
$\langle \alpha^0|H_{\rm D}|\beta^0\rangle \sim \delta_{\alpha,\beta}$ in every degenerate subspace.
This ensures that the corrected wave function does not diverge due to singularities.
We focus on the ground-state subspace, but the procedure is identical for any subspace.
Given $\kk$ ground states $|\n_{\text{gs}}^0 \rangle$ of ${\cal H}_{\rm P}$ with energy $E_{\text{gs}}^0$, we need to form the $\kk \times \kk$ subspace matrix $V_{\n,\m}=\langle \n_{\text{gs}}^0|H_{\rm D}|\m_{\text{gs}}^0\rangle$; as shown in Fig. \ref{FigVD}.
Because $H_{\rm D}$ is Hermitian, $V$ is too.
Every Hermitian matrix can be diagonalized by a unitary transformation ($\U^{-1} \V \U=\D$) and
we find the correct linear combinations $|\alpha_{\text{gs}}^0\rangle$ in the columns of $\U$.
It satisfies $\langle \alpha_{\text{gs}}^0|V|\beta_{\text{gs}}^0\rangle \sim \delta_{\alpha,\beta}$ since D is diagonal.
The diagonal entries of $D$ are the eigenvalues of $V$ and also the first order
energy corrections $E_\alpha^1$. We need to pick the lowest eigenvalue $E_{\alpha,\text{low}}^1$
and find the corrected ground state energy $E_{GS}=E_{\text{GS}}^0+E_{\alpha,\text{low}}^1$.
The corresponding $l$ eigenvectors $|\alpha_{\text{gs}}^0\rangle$
will now determine the sampling behavior, since the annealing state will be in
their span.
The following scenarios can occur:

(i) {$\lv=1$}: In this case $\p_i=\langle n_{\text{gs}}^0 |\phi_{\alpha,\text{low}}^0\rangle^2$,
because there is a single state $|\phi_{\alpha,\text{low}}^0\rangle$.
If sampling is fair, it will remain fair, regardless of how much the
higher energy eigenvalues of ${\cal H}_{\rm P}$ change during the
adiabatic anneal.  If certain states have $\p_i = 0$, $|n_{\text{gs}}^0 \rangle$ will
never be available at the end of the anneal.

(ii) {$\l>1$}: Let $A$ be the $k \times m$ matrix consisting
of all $l$ $|\phi_{\alpha,\text{low}}^0\rangle$. If there is a vector $x$ such that
$A x=y$ and $y_i \cdot y_i^*=1$ for all $i$, then fair sampling is
potentially possible according to first order. If there exists an $i$ such that $y_i = 0$ for all $x$,
then that ground-state is never found. The same argument can be made for
biased sampling where there is no suppression but certain states are
over-sampled.

\begin{figure}[h!]
\includegraphics[width=1\columnwidth]{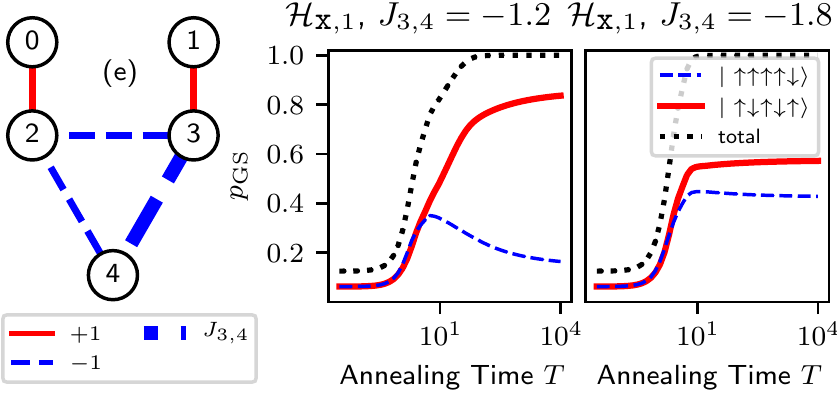}
\caption{\label{Fig2}
Toy problem with $\N = 5$ variables and both ferromagnetic
(solid lines) and antiferromagnetic (dashed lines) interactions of
different strength (thickness of the lines) integrated using
\textsc{qutip} \cite{johansson:13}. Data for a driver
${\cal H}_{\X,1}$. By changing $J_{3,4}$ one can change
the sampling bias arbitrarily.}
\end{figure}

(iii) {$\V$ is zero}: All eigenvalues $E_\alpha^1=0$ and
the sampling probabilities are determined by second-order perturbation,
i.e., the probabilities depend on higher eigenvalues of ${\cal H}_{\rm
P}$ (see Fig.~\ref{Fig2}).

The second-order perturbation terms only play a relevant role if $\V$ is
trivial. If $\l>0$, the sampling behavior is determined by $\V$ which does
not depend on ${\cal H}_{\rm P}$. This means that the sampling behavior
is purely a property of the driver Hamiltonian ${\cal H}_{\X,n}$ and the
ground-state eigenvectors of ${\cal H}_{\rm P}$. We have verified this
on numerous small systems, as well as structured and random-coupling
systems with direct integration and were always able to predict the
sampling probabilities that correspond to the state found after the
anneal.

Figure \ref{Fig1}(a) is the example studied in Ref.~\cite{matsuda:09},
where ${\cal H}_{\X,2}$ leads to fair sampling.  There are six degenerate
ground states, two of which are suppressed.  With a driver of the form
${\cal H}_{\X,1}$ we obtain $\l>1$, meaning that there are multiple
$|\alpha_{\text{gs}}^0\rangle$ states that determine the sampling. However,
the suppressed states have $\p_i = 0$.  In Fig.~\ref{Fig1}(b) we show a
more complex example -- the smallest problem we were able to find that
has $\l=1$ and one state where $\p_i = 0$. It is a $12$-fold degenerate
system with two states fully suppressed when ${\cal H}_{\X,1}$ is used
as a driver.  The fact that $\l=1$ could be a reason why the suppression
sets in earlier during the anneal. This case is problematic for
annealing schedules that are fast quenches, because there is a much
smaller window during the anneal where the total ground-state
probability is approximately unity and the suppressed state has not yet
reached zero probability.  Using ${\cal H}_{\X,2}$ results in fair
sampling.  Figure \ref{Fig1}(c) shows a system that has six ground
states with two ground states in hard suppression.  Using ${\cal
H}_{\X,1}$ as a driver, we obtain $\l=1$ and a unique
$|\alpha_{\text{gs}}^0\rangle$ with two hard suppressed states with zero
probability. For ${\cal H}_{\X,2}$, $\l > 1$ we obtain multiple
$|\alpha_{\text{gs}}^0\rangle$ states. However, two states are hard
suppressed. Using ${\cal H}_{\X,3}$ as a driver results in $\l=1$ and a
unique $|\alpha_{\text{gs}}^0\rangle$. However, there is a soft suppression of
two ground states (not shown). Finally, using ${\cal H}_{\X,4}$ we
obtain $\l = 1$ and fair sampling. The case shown in Fig.~\ref{Fig1}(d)
reveals the undersampled states when ${\cal H}_{\X,1}$ is replaced by
${\cal H}_{\X,2}$. More precisely, it changes from $\l>1$ with four soft
suppressed states to $\l = 1$ with the previously two oversampled state now
being undersampled.  Using a driver ${\cal H}_{\X,3}$ results in $\l = 1$
and fair sampling.

Figure \ref{Fig2} shows a problem where by changing the strength
of $\J_{3,4}$ one can change the sampling bias arbitrarily. Note that
changing $\J_{3,4} = -1.2$ to $-1.8$ does shift the
relative energies of the ground state and the various low excited states,
but does not change their order. In terms of perturbation theory, $\V$
is trivial, and second-order perturbations dictate the behavior of
the system. Because there are terms $\propto 1/(E_i - E_{\rm GS})$ with
$E_{\rm GS}$ the ground-state energy, shifting the energy levels $E_i$ will
influence the sampling.

\begin{figure}[h!]
\includegraphics[width=1\columnwidth]{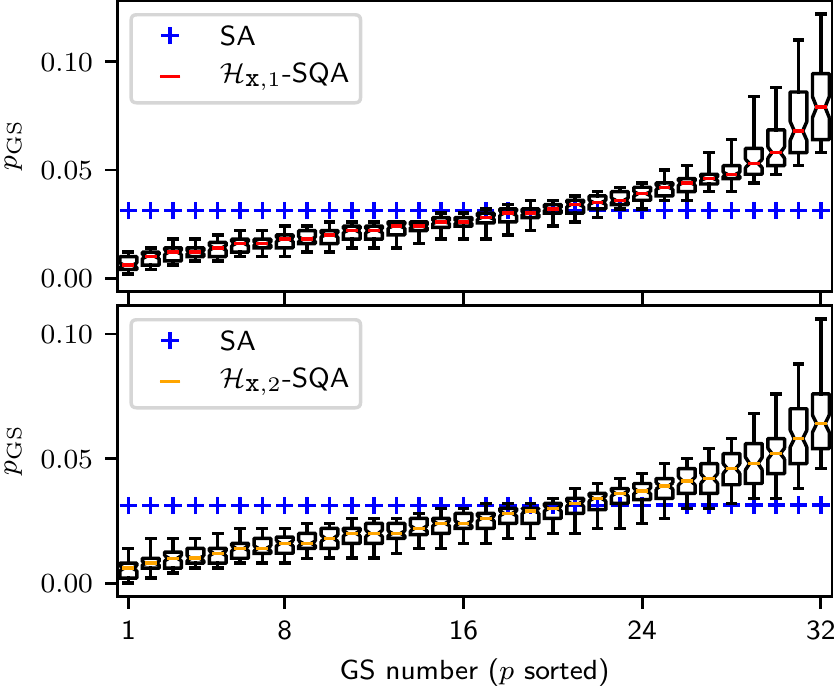}
\caption{\label{Fig4}
Rank-ordered probability $\p_{\rm GS}$ to find different degenerate
states for a two-dimensional Ising spin glass with $\N = 8^2 = 64$
and a ground-state degeneracy of $32$. The data are averaged
over $100$ disorder realizations. For each instances, $500$ independent
runs are performed and the probability to find a given ground-state
configuration computed. While simulated annealing (SA) samples close to
fair, both SQA-${\cal H}_{\X,1}$ and SQA-${\cal H}_{\X,2}$ show a
clear bias in the sampling. In particular, there is no notable
improvement of using a driver of the form ${\cal H}_{\X,2}$ over a
transverse-field driver ${\cal H}_{\X,1}$. We have also simulated
systems with up to $\N = 12^2 = 144$ variables and ground states with
up to $96$-fold degeneracy obtaining similar results. Note that the bias
becomes more pronounced for increasing system size $\N$ (not shown).}
\end{figure}

{\em Quantum Monte Carlo results}. To corroborate our results with
larger systems, we perform a fair-sampling study analogous to the one
done in Ref.~\cite{mandra:17} for two-dimensional Ising spin glasses on
a square lattice with periodic boundary conditions. The couplers are
chosen from $\J_{i,j} \in \{\pm 1, \pm 2,\pm 4\}$. This ensures that
degeneracies are small.  The coupler-configuration space is mined for
specific degeneracies as done in Ref.~\cite{mandra:17}. Figure
\ref{Fig4} shows representative rank-ordered probabilities to find
different minimizing configurations using simulated annealing (SA)
\cite{kirkpatrick:83,isakov:15,moreno:03}, as well as transverse-field
simulated quantum annealing (SQA-${\cal H}_{\X,1}$)
\cite{santoro:02,heim:15,isakov:16,mazzola:17,koenz:18x,comment:x_sqa_param}
and simulated quantum annealing with a \emph{stoquastic} two-spin
driver (SQA-${\cal H}_{\X,2})$ \cite{comment:xx_sqa_param}
\cite{mazzola:17a}. The data are averaged over $100$ disorder
realizations. While the data for SA for this particular problem show a
fair sampling of all minimizing configurations, neither a
transverse-field ${\cal H}_{\X,1}$ nor a more complex ${\cal H}_{\X,2}$
driver can remove the bias. This suggests that even if QA machines with
more complex ${\cal H}_{\X,2}$ drivers are constructed, sampling will
remain unfair unless post-processing is applied \cite{ochoa:18x}. The
close connection between SQA and QA performances is discussed in
Refs.~\cite{isakov:16,mazzola:17}.

{\em Effects of more complex drivers}. The following section shows
that any driver (stoquastic or non-stoquastic) needs to be sufficiently dense to sample fair for generic
larger systems. To predict the sampling probabilities it is sufficient
to know $\V$ (except when $\V$ is trivial). $\V$ can be constructed with
only the ground-state eigenvectors of ${\cal H}_{\rm P}$ (no
eigen energies needed) and the driver ${\cal H}_{\X,n}$.  This can be
used to analyze different drivers---without specifying a concrete
problem Hamiltonian ${\cal H}_{\rm P}$---by merely sampling from
possible ground-state combinations. As an example consider a twofold
degeneracy in a five-spin system. Because we want to test the driver for
all possible ground-state combinations, we can exhaustively generate all
the ground-state pairs, i.e., $\N(\N-1)/2$, where $\N=2^5$ and check $\V$
for each one pair, to analyze the sampling behavior. For larger $N$, we
sample instead of searching exhaustively.

\begin{figure}
\includegraphics[width=1\columnwidth]{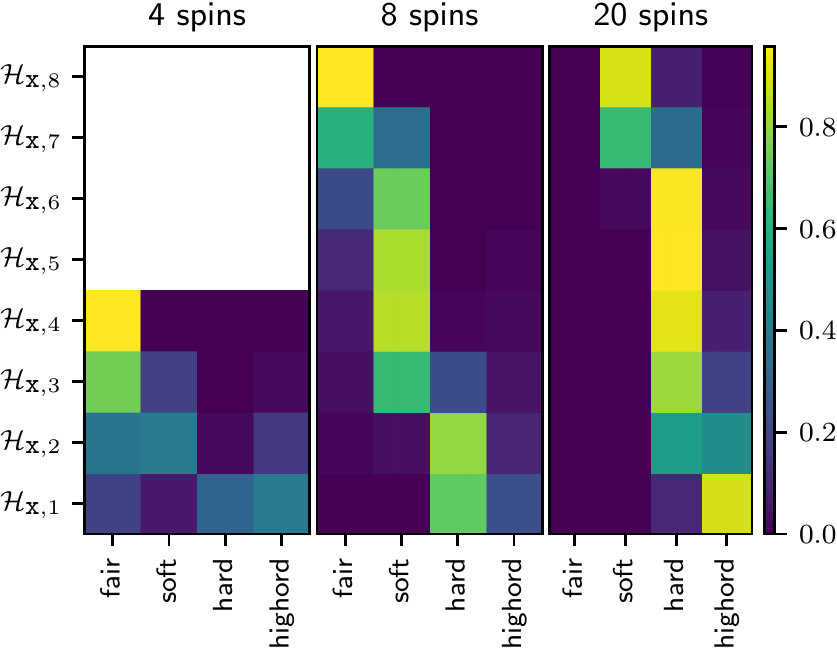}
\caption{\label{Fig3}
For each spin system all possible degeneracies are sampled with $400$
random ground-state combinations.  For small systems, all combinatorial
possibilities of ground states can be checked, and the random sampling
approximates the exact result fast.  For four spins, fair sampling is
reached for all possible problem Hamiltonians ${\cal H}_{\rm P}$, once
the driver matrix is dense, i.e., all off-diagonal contain nonzero
entries,.  For example, this is the case when using ${\cal H}_{\X,4}$
for a system with four spins.  Similarly, for eight spins the system moves
from hard to soft, to fair sampling as more complex drivers are used.
As the system with $20$ spins illustrates, below for any $\n < 7$ in
${\cal H}_{\X,n}$ there is only a dependence on second-order
perturbation (see Fig. \ref{Fig2}) or hard suppression in the average
case.}
\end{figure}

Figure \ref{Fig3} shows how probable it is for a random degeneracy and a
ground-state combination to be sampled according to the following
categories:

{\it fair}: All ground states have the same probability.

{\it soft}: At least one ground state is soft suppressed with a
ratio smaller than $1:100$ (least likely vs most likely).

{\it hard}: At least one ground state is soft suppressed with a
ratio larger than $1:100$ or not found at all, i.e., hard suppression.
For better visibility in Fig.~\ref{Fig3} we combine these two cases.
However, most of the time the suppression is hard.

{\it highord}: The matrix $\V$ is trivial. Higher-order
perturbation will determine the sampling behavior.  In the
generic case of random couplings this leads to both soft or hard
suppression.

In all cases and for ${\cal H}_{\X,n}$ with $\n \le 8$ we use
$\Gamma^{\X,n} = 1$. Using different values for the different amplitudes
leads to worse sampling, because the matrix $\V$ has multiple
different entries. A random matrix has a unique eigenvector which is
not parallel to $(1, 1,\ldots, 1, 1)$ in the generic case.  Hence,
introducing more variety into $\V$ leads to more unique (and unfair)
$|\alpha_{\text{gs}}^0\rangle$. How the ratio of soft to hard suppression is influenced
by this was not investigated, since it is unfair in the generic case.
Repeating multiple annealing runs with individually randomized
$\Gamma^{\X,n}_{i,j,...}$ and averaging improves the sampling but, if
not dense enough, will not be able to remove all hard suppression in a
generic case.

{\em Conclusions}. We have studied the necessary ingredients needed
for quantum annealing to sample ground states fairly. From
Fig.~\ref{Fig3} we surmise that a fairly dense driver is needed to
obtain fair sampling.  Carefully controlling the anneal with additional
parameters, for example as shown in Ref.~\cite{susa:18} might help
mitigate the bias, however, this remains to be tested. We do emphasize,
however, that a ${\cal H}_{\X,2}$ driver with the typical annealing
{\it modus operandi} used in current hardware will not yield a fair sampling
of states and performs comparably to an elementary transverse-field driver ${\cal H}_{\X,1}$.

\begin{acknowledgments}

M.~S.~K.~thanks Dominik Gresch for insightful discussions that
accelerated the discovery process. The large-scale simulated quantum
annealing calculation were done using the Monch Cluster at ETH Zurich.
H.~G.~K.~would like to thank Salvatore Mandr{\`a} for multiple
discussions and acknowledges the ARC Centre for Excellence in All-Sky
Astrophysics in 3D (ASTRO 3D) East Coast Writing Retreat for
support in preparing the manuscript.  H.~G.~K.~acknowledges support from
the NSF (Grant No.~DMR-1151387).  H.~G.~K.'s research is based upon work
supported by the Office of the Director of National Intelligence (ODNI),
Intelligence Advanced Research Projects Activity (IARPA), via
Interagency Umbrella Agreement No.~IA1-1198. The views and conclusions
contained herein are those of the authors and should not be interpreted
as necessarily representing the official policies or endorsements,
either expressed or implied, of the ODNI, IARPA, or the U.S.~Government.
The U.S.~Government is authorized to reproduce and distribute reprints
for Governmental purposes notwithstanding any copyright annotation
thereon.  We thank the Texas Advanced Computing Center (TACC) at The
University of Texas at Austin for providing HPC resources (Stampede
Cluster) and Texas A\&M University for access to their Ada, and Lonestar
clusters.

\end{acknowledgments}

\bibliography{refs,comments}

\end{document}